\documentclass[preprint]{iucr}              
\usepackage{graphicx}
\usepackage{amsmath}

     \journalcode{J}              

\begin{document}                  



\title{The \textit{pypadf} package: computing the pair-angle distribution function from fluctuation scattering data}


\cauthor[a]{Andrew V. }{Martin}{andrew.martin@rmit.edu.au}{}
\author[a]{Patrick}{Adams}
\author[a]{Jack}{Binns}


\aff[a]{School of Science, STEM College, RMIT University, Melbourne, Victoria  3000 \country{Australia}}









\maketitle                        


\begin{abstract}
The pair-angle distribution function (PADF) is a three- and four-atom correlation function that can characterise the local angular structure of disordered materials, particles or nanocrystalline materials. The PADF can be measured by x-ray or electron fluctuation diffraction experiments, which can be collected by scanning a small beam across a structurally disordered sample or flowing a sample across the beam path. It is a natural generalisation of the established pair-distribution methods, which do not provide angular information. This software package provides tools to calculate the PADF from from fluctuation diffraction data. The package includes tools for calculating the intensity correlation function, which is a necessary step in the PADF calculation and also the basis for other fluctuation scattering techniques.
\end{abstract}


\section{Introduction}
\;\\
Fluctuation scattering techniques have been developed for studying the  structure of disordered materials such as colloidal materials, liquid crystals and amorphous solids \cite{Treacy2005, Kurta2016,Zaluzhnyy2019}, and for single particle imaging of, for example, proteins, viruses, and nanoparticles \cite{Kirian2012}. Depending on the context in which they were developed, these techniques have a variety of names including fluctuation electron microscopy ~\cite{Treacy2005,Fan2005}, fluctuation x-ray microscopy, fluctuation scattering ~\cite{Kam1977,Kam1981,Saldin2009,Saldin2010}, and x-ray cross-correlation analysis ~\cite{Wochner2009,Kurta2016,Zaluzhnyy2019,Lehmkuehler2014}. These methods all use statistical approaches to extract structural information from a set of diffraction measurements (10$^2$--10$^5$) where the sample structure and/or orientation varies randomly between measurements. Historically, the applications to imaging and to disordered materials have developed in parallel, because the type of structural information extracted is different. In the imaging applications, the goal is to recover a 3D image of a reproducible particle, whereas for disordered materials the goal is to probe some characteristic local 3D structure.

In applications to disordered materials, the sample structure is assumed to lack long-range order so that moving the beam to a new area of the sample generates statistical fluctuations in the scattering~\cite{Treacy2005,Fan2005}. The statistical properties such as the variance \cite{Treacy2005} or angular symmetries~\cite{Wochner2009} are then obtained from the ensemble of diffraction measurements. Many fluctuation techniques compute an angular intensity cross-correlation function, which captures both intensity variance and intensity cross-correlations as a function of angular separation~\cite{Kam1977,Kurta2016}. The correlation function is then used to distinguish structural models, identify the presence of symmetric local structures, map local structures or extract the relative composition of ideal structural models. These methods have been applied to colloidal particles~\cite{Wochner2009,Liu2017,Liu2022}, nanoparticles~\cite{Niozu202} and liquid crystals~\cite{Zaluzhnyy2015,Martin2020a}, texture in polycrystalline materials~\cite{Binns2022} and metallic glasses\cite{Liu2013}. Many of these experiments observe trends correlations as a function of space, temperature or sample composition. Alternatively, correlation functions have been matched to models of local structure to extract information about the distribution of local structures. Despite this progress, the structural interpretation of correlation-based analysis results remains an outstanding issue for the field. 



To address the challenge of obtaining interpretable local 3D structural information from a disordered sample, the pair angle distribution function (PADF) technique was developed~\cite{Martin2017}. The PADF is a real-space three- and four-atom distribution that can be extracted from fluctuation scattering data by applying a linear transformation to the angular intensity cross-correlation functions. It applies to local 3D structure in bulk disordered materials or, potentially, to heterogeneous, single particles in uniformly random orientations. The PADF provides information about two pair distances and the relative angle between the two pairs. It contains bond-angle information and other local angular structure that can be used to ``fingerprint" the local atomic arrangements in a disordered material. It has been applied with x-rays to identify local structures in lipid cubic phase~\cite{Martin2020a}, with electrons to study defects in disordered carbons~\cite{Martin2020} and there are prospects for studying proteins~\cite{Adams2020} and colloids~\cite{Bojesen2020}. The PADF is primarily designed for fluctuation studies of disordered materials because these are the cases where it remains difficult to obtain structural insights from the analysis of the correlation function in $q$-space. In principle, the PADF can be calculated from single particle fluctuation data, but provides a less direct representation of structure than a 3D imaging via correlations~\cite{Donatelli2015}. 

Here we present a python package for the calculation of the PADF from diffraction data. The code provides a tools to i) calculate a $q$-space correlation volume from a fluctuation scattering dataset, ii) applying masking and geometric corrections to the correlation volume, iii) calculate the PADF from the correlation volume, and iv) plotting tools for the correlation volumes and the PADF volumes.

\section{Overview of the \textit{pypadf} package}
\;\\
The \textit{pypadf} software is written in \textit{Python 3} and provides tools for computing angular correlation functions and PADFs from experimental or simulated diffraction data. These tools are intended for use on data for a wide variety of facilities including synchrotrons, X-ray free-electron lasers and electron microscopes.  By default it processes data stored as one diffraction pattern per file. However, the \textit{pypadf} package has been designed in a modular fashion to enable future extensions to the file formats of different facilities as required.

\begin{figure}
\begin{flushleft}
\includegraphics[angle=90, width=6.5cm]{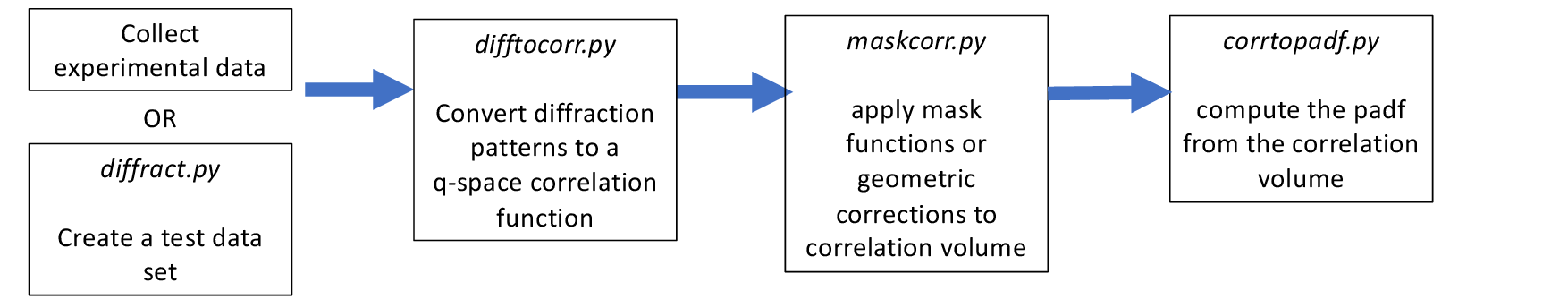}
\end{flushleft}
\caption{A schematic of the workflow for computing the $q$-space correlation function and the PADF. The \textit{pypadf} package consists of separate scripts that are run in the order indicated by the arrows. 
[NOTE: fix inconsistent capitalisation in text boxes]
}
\label{fig:padfworkflow}
\end{figure}

The \textit{pypadf} package has three parts: (i) the main scripts, (ii) the \textit{params} module containing all input parameter specifications and tools and (iii) the \textit{fxstools} module that contains the functional tools for loading, analysing and plotting correlation data. As shown in Fig. \ref{fig:padfworkflow}, the main scripts are \textit{difftocorr.py} to convert diffraction patterns to a correlation function, \textit{maskcorr.py} to prepare the correlation function for the PADF calculation and \textit{corrtopadf.py} that computes the PADF from the correlation function. The correlation and PADF functions can be plotted with the script \textit{plotfxs3d.py}.  Each script imports a submodule from the \textit{params} module that defines parameters specific to that script. The scripts take input parameters from a configuration file or via command line options. The configuration file is read first and then command line arguments are read second. Hence, parameter values from command line arguments take precedence (i.e. will override) parameter values defined in the configuration file. This enables the commandline options to be used in batch scripts where a small number of parameters are changing for each dataset. The configuration file contains parameter values that remain constant, while the command line options can be used to conveniently update parameters that are changing between datasets. Example configuration files are included with the code. In the rest of this section we detail the main scripts for computing the PADF including diffraction simulation, correlation calculation and finally the PADF calculation. For each step we outline background theory, the numerical implementation of the equations and the script associated with each step.

The \textit{pypadf} package includes scripts for basic tools for simulating elastic scattering that are summarised in Appendix \ref{Appendix:Diffraction}, which are provided for convenient testing of the correlation and PADF scripts. We note that for detailed simulation studies there are established diffraction programs available for both single particles and crystals such as \textit{Reborn}~\cite{Kirian2020,Chen2021}, \textit{Condor}~\cite{Hantke2016} and \textit{MLFSOM}~\cite{Holton2014}.




\subsection{The Angular Intensity Correlation Function $C(q,q',\theta)$}
\label{sec:corr}

\subsubsection{Intensity correlations: mathematical and numerical details}
\;\\

The angular intensity correlation function is calculated from the polar representations of the diffraction data $I(q,\phi)$ as follows
\begin{equation}
C(q,q',\Delta\theta) = \frac{1}{N}\sum_i^N \int I(q,\phi') I(q',\phi' + \Delta\phi) d\phi' \;.
\label{eq:Cqq}
\end{equation}
where $N$ is the number of diffraction patterns in the dataset.
We assume that the diffraction patterns, $I(q,\phi)$, used to compute the q-space correlation function have been corrected for any solid angle and polarisation effects.  

In the polar representation, each $q$ and $q'$ value labels an intensity ring. Numerically, the 1D fast Fourier transform and the convolution theorem are used to compute the angular correlation between each pair of rings $q$ and $q'$.

The correlation background due to static signals can be estimated from the cross-correlation of independent diffraction patterns. Such static signals include background scattering in the measured images that does not vary from frame to frame. The effect of static background signals can be estimated  by randomly correlating pairs of diffraction patterns
\begin{equation}
C_{BG}(q,q',\Delta\theta) = \frac{1}{N}\sum_i^N \int I_i(q,\phi') I_{j(i)}(q',\phi' + \Delta\phi) d\phi'  \;.
\label{eq:Cbg}
\end{equation}
where $j(i)$ is a randomly chosen index that is not equal to $i$. This background estimate can be subtracted from the estimate of the correlation signal made by Eq. (\ref{eq:Cqq}).

A background subtracted correlation function can be computed in a single pass over the data by computing difference correlations~\cite{Mendez2016}
\begin{align}
C_{DIFF}(q,q',\Delta\theta) &= \frac{1}{N}\sum_i^N \int  \Delta I_{i,j(i)}(q,\phi') \Delta I_{i,j(i)}(q',\phi' + \Delta\phi) d\phi'  \;, 
\label{eq:Cdiff}
\end{align}
where $\Delta I_{i,j(i)}(q,\phi')  = I_i(q,\phi') - I_{j(i)}(q',\phi')$ for pairs of randomly picked diffraction patterns $i \neq j(i)$. It can be shown that $C_{DIFF} = 2(C - C_{BG})$.

A mask can be applied to exclude the beamstop, detector gaps and bad pixels from analysis. In the \textit{pypadf} package, a binary mask is used that takes a value 1 for included pixels and 0 for excluded pixels. The effect of the mask on the correlation function is corrected by dividing the correlation function by the correlation of the mask: $C_{corrected}(q,q',\theta) = C(q,q',\theta) / C_{mask}(q,q',\theta)$ wherever $C_{mask}(q,q',\theta)>0$. Where the correlation of the mask is 0, the corrected correlation function is set to zero.


\subsubsection{\textit{difftocorr.py} : computing correlations from diffraction patterns}
\;\\

The script \textit{difftocorr.py} computes a correlation function from a set of diffraction patterns. It can compute $C(q,q',\Delta\theta)$, $C_{BG}(q,q',\Delta\theta)$ or $C_{DIFF}(q,q',\Delta\theta)$, and it can perform the mask correction. By default it calculates two correlation functions from the odd and even frames, which can be compared to visually check that the correlation functions have converged. The script assumes that diffraction patterns are saved in individual files and it constructs a filelist from the folder based on a filename format specified in the configuration file. There are parameters to centre, crop, rebin and mask the diffraction patterns. The processed diffraction patterns can be saved to check that the processing parameters are correct. The script requires detector geometry parameters including the sample-to-detector distance, the beam center, the width of a detector pixel and the wavelength. The output of the script is the 3D correlation function saved in a numpy file or as a raw binary file.


\subsubsection{\textit{maskcorr.py} : applying corrections to the correlation volume}
\;\\

The script \textit{maskcorr.py} makes modifications to the correlation volume prior to computing the PADF. Accurate calculation of the PADF requires the correlation volume to be evenly sampled with respect to $\cos\theta$. However, the correlation function is most conveniently calculated with uniform $\theta$ sampling. A multiplication by $|\sin\theta|$ corrects for the sampling.

In experiment, it can occur that the effects of background scattering or other artefacts are still evident after calculating $C_{DIFF}$ or subtracting $C_{BG}$. If these spurious signals are confined to a particular region of the correlation function, such as low $q$ values, they can be masked. The script $\textit{maskcorr.py}$ can apply low and high pass filters on the q dimensions. It can also apply a mask to the region close to $\theta = 0$, which can remove correlation artefacts due to noise on the diffraction patterns. The noise may be detector noise or noise from spurious correlations in the sample, e.g. across distances larger than the structural correlation length in the sample or coherent interference between distant atoms.

\subsection{Pair-angle distribution function - $\Theta(r,r',\theta)$}

\subsubsection{The PADF: mathematical and numerical details}
\;\\
A principal purpose for \textit{pypadf} is to provide code to convert the correlation function $C(q,q',\theta)$ into the pair-angle distribution function (PADF), denoted $\Theta(r,r',\theta)$. Here we review the transformation of the $C(q,q',\theta)$ into real-space and the definition of the PADF.

The sample's scattering factor can be expanded in terms of spherical harmonics as follows
\begin{equation}
|F_m(\textbf{q})|^2 = \sum_{lm} I_{lm}(q) Y_{lm}(\theta,\phi) \;,
\end{equation}
where $Y_{lm}(\theta,\phi)$ are spherical harmonic functions (we can assume real spherical harmonics). It can be shown that the correlation function, has the form
\begin{equation}
C(q,q',\theta) = \sum_l P_l\left(\frac{\textbf{q}\cdot\textbf{q}'}{|\textbf{q}||\textbf{q}'|}\right) B_l(q,q') \;,
\label{eq:CBrelation}
\end{equation}
where $P_l(x)$ are the Legendre polynomials. The $B_l(q,q')$ matrices are given by
\begin{equation}
B_l(q,q') = \sum_m I_{lm}(q) I_{lm}(q') \;.
\label{eq:Blqq_Ilm}
\end{equation}
The $B_l(q,q')$ matrices can be extracted by numerically inverting Eq. (\ref{eq:CBrelation}) using singular value decomposition (SVD) or by using the orthogonality properties of the Legendre polynomials. The default behaviour in the \textit{pypadf} package is to use SVD. A value of 0.5 is used to regularize the small singular values, which was selected to exclude singular values to near 0. The singular values depend on the experiment geometry, via the Ewald sphere, but not on the input data. The value of 0.5 has been adequate for all applications and tests that we have made to date.

The $B_l(q,q')$ are converted to real-space using two spherical Bessel transforms:
\begin{equation}
B_l(r,r') \equiv  (-1)^l 16 \pi^2  \int\int  q^2 q'^2 j_{l}(qr) j_{l}(q'r') B_l(q,q') dq dq' \;,
\label{eq:sphB_transform}
\end{equation}
The spherical Bessel transform is implemented using the discrete form of Lanusse et al.~\cite{Lanusse2012}. In the discrete form, a general function $f(q)$ of the radial $q$ coordinate is transformed to a real space function by
\begin{equation}
f(r) = \sum^{\infty}_{n=1}  \frac{\sqrt{2\pi} R^{-3}}{j^2_{l+1}(q_{nl})} j_l\left(\frac{q_{ln} r}{R}\right) f_l\left(\frac{q_{ln}}{R}\right) \;,
\label{eq:sphB_discrete}
\end{equation}
where $R$ is the maximum value of $r$ and $q_{ln}$ is the $n^{th}$ zero of the $j^{th}$ spherical Bessel function. To implement in this, all $B_l(q,q')$ matrices are first computed on the zero positions of the $l=0$ spherical Bessel function. Interpolation is then use to remap matrices with $l>0$ onto the appropriate $q$ sampling points before using Eqs. \ref{eq:sphB_discrete} and  to compute the real-space $B_l(r,r')$ matrices.

The code computes a real-space correlation function by forming a weighted sum of the $B_l(r,r')$ matrices:
\begin{equation}
C(r,r',\cos\theta) = \sum_l P_l\left(\frac{\textbf{r}\cdot\textbf{r}'}{|\textbf{r}||\textbf{r}'|}\right) B_l(r,r') \;.
\end{equation}
The function $C(r,r',\cos\theta)$ is a scaled form of the pair-angle distribution function $\Theta(r,r',\theta)$ as follows
\begin{equation}
C(r,r',\cos\theta) = \frac{\rho^4_0 I_0^2}{r r' \sin\theta} \Theta(r,r',\theta) \;.
\label{eq:Crr_to_PADF}
\end{equation}
where $I_0 = r_e^2 \frac{ \phi_0} {A^2}$, recalling that solid angle and polarization effects are already assumed to be corrected.

\begin{figure}
\includegraphics[scale=0.4, angle=270]{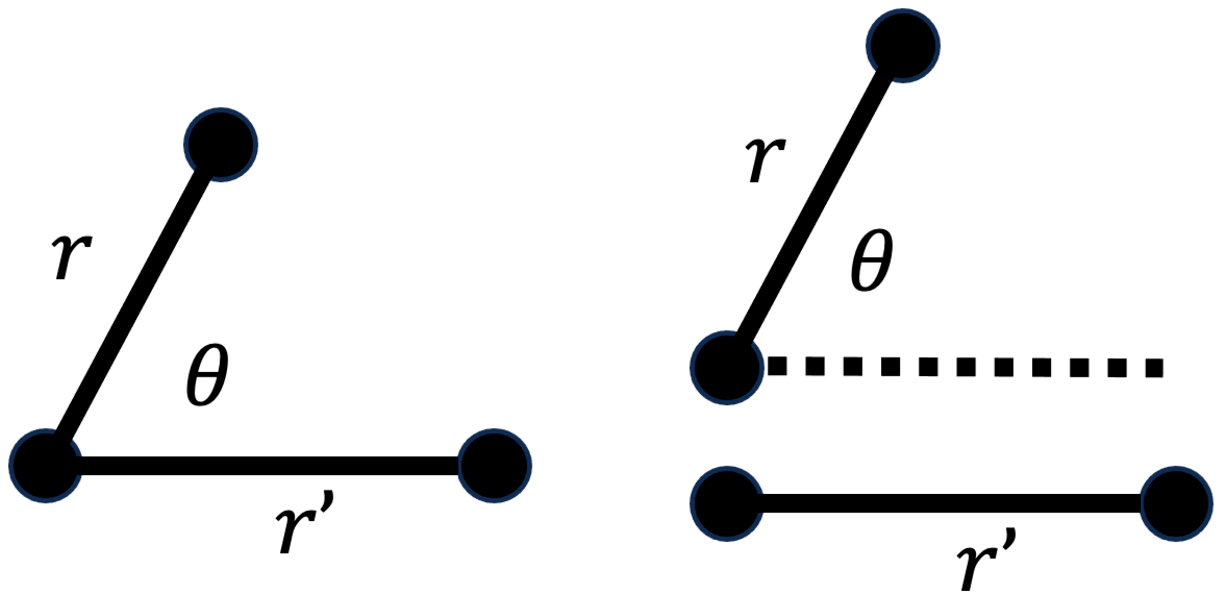}
\caption{The relevant coordinates of 3-atom combinations (left) and 4-atom combinations (right) that contribute to the PADF. The PADF is not sensitive to absolute position, absolute orientation or, in the 4-atom case, the separation distance between the two atom pairs.
}
\label{fig:padfcoordinates}
\end{figure}

It can be shown that the PADF can be written as
\begin{equation}
\Theta(r,r',\theta) = \int \int g^{(2)}(\textbf{r}) g^{(2)}(\textbf{r}') \delta( \cos\theta - \hat{\textbf{r}} \cdot \hat{\textbf{r}}') d\Omega_r d\Omega_{r'} \;.
\end{equation}
or
\begin{equation}
\Theta(r,r',\theta) = \tilde{g}^{(2)}(r,r,0) + \tilde{g}^{(3)}(r,r',\theta) + \tilde{g}^{(3)}(r,r',\pi-\theta) + \tilde{g}^{(4)}(r,r',\theta) \;.
\end{equation}
The functions $\tilde{g}^{(n)}(r,r,\theta)$ are multi-atom correlation functions, parametrised by two pair distances and a relative local angle. The tilde symbol indicates that these terms differ from the general correlation functions from statistical mechanics by integrating out the degrees of freedom that the diffraction is insensitive to, such as the absolute position and absolute orientation of the pairs. The distance between the pairs is also integrated out. The remaining degrees of freedom are shown in the diagram in Fig. \ref{fig:padfcoordinates}.


\subsubsection{\textit{corrtopadf.py} : computing the PADF from the correlation function}
\;\\
The script \textit{corrtopadf.py} converts the q-space correlation function into the PADF. It calculates the $B_l(q,q')$ matrices, applies the numerical spherical Bessel transforms to obtain $B_l(r,r')$ matrices, then reconstructs the PADF. The $B_l(q,q')$ and $B_l(r,r')$ matrices can be saved as optional output. The number of spherical harmonics is set to control the angular resolution and only even spherical harmonics are used, because the inclusion of odd harmonics reduces the accuracy of the matrix inversion. Neglecting absorption, the approximation to remove odd harmonics becomes valid. The default behaviour is to compute $C(r,r',\cos\theta)$ because it requires less knowledge from the user. As per Eq. \eqref{eq:Crr_to_PADF}, the output $C(r,r',\cos\theta)$ can be multiplied by $r r' \sin\theta$ to produce a function proportional to the PADF.  There is an option to multiply by the constants in Eq. \eqref{eq:Crr_to_PADF} to obtain absolute values of the PADF.




\section{An example PADF calculation}
\begin{figure}
\begin{center}
\includegraphics[scale=0.5]{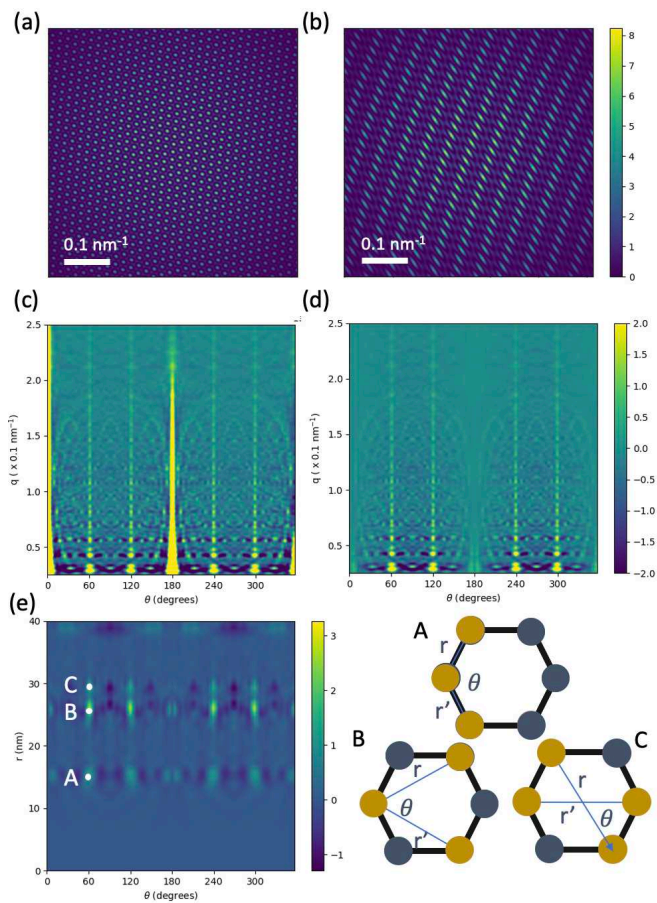}
\end{center}
\caption{(a), (b) Two diffraction patterns from the model hexagonal sample. (c) The $q$-space correlation function for the hexagonal model calculated from 1000 patterns without the $r r' \sin\theta$ correction. (d) The $q$-space correlation function from (c) with the $r r' \sin\theta$ correction. (e) PADF of the hexagonal structure computed from the corrected $q$-space correlation function. 
[Note: double check the scale bar in (a), (b)] 
}
\label{fig:hexexample}
\end{figure}

\;\\ 
Here we provide an example of a PADF calculated from a simulated set of fluctuation scattering diffraction patterns. The model sample contains six point scatterers in a hexagonal arrangement with a nearest neighbour distance of 15 nm. A dataset of 1000 diffraction patterns were simulated with \textit{diffract.py}. For each pattern the diffraction pattern was rotated to a random orientation. The maximum $q$ value recorded at the edge of the detector is 0.25 nm$^{-1}$, which corresponds to a resolution of 4 nm. At this resolution, the atomic scattering factors are relatively flat and the point scatterers could be represented by any atom. Example diffraction patterns from the dataset are shown in Fig. \ref{fig:hexexample}(a) and (b). The regularity of the observed interference patterns arises from the geometric arrangement of the six scatterers and their absolute orientation. There is a weak attenuation at high q due to the atomic scattering factor and the reduced solid-angle of pixels near the edge of the detector. No noise is modelled on the detector.

The correlation function computed from all 1000 simulated patterns is shown in Fig. (c). Since no background signals have been modelled, the standard correlation function $C(q,q',\theta)$ defined by Eq. (\ref{eq:Cqq}) has been calculated. A highest $l$ value in the spherical harmonic expansion was 30, which corresponds to an angular resolution of 12 degrees.

There are strong features at angles of 60 and 120 degrees, which is expected from the hexagonal arrangement of atoms in the sample. The correlation function is strongest at 0 and 180 degrees, which is expected because the diffraction pattern is sampled with an even sampling in $\theta$. Fig \ref{fig:hexexample}(d) shows the $\sin\theta$ correlation after applying a $\sin\theta$ scaling to convert to an even $\cos\theta$ sampling [see Eq. \eqref{eq:Crr_to_PADF}], as this is a necessary prior step for accurate PADF measurements. The PADF was calculated using \textit{corrtopadf.py} and the result shows the expected angular peaks for a hexagonal arrangement of scatterers at $r=r'=$ 15 nm, 26 nm and 30 nm, and at 60 and 120 degrees. The ratios of the peak heights are within 5\% of the ideal peak ratio values as shown in Table \ref{table:hexpeaks}. This shows that both the peak positions and the peak heights can be analysed quantitatively. As mentioned above, the raw output of \textit{corrtopadf.py} is multipled by $|\sin\theta|$ to generate the PADF and this is necessary for analysing peak heights. However this multiplication is inconsistent with the finite number of angular basis functions and artificially lowers the values near $\theta=0$ . Hence, peak heights near $\theta=0$ can not be analysed quantitatively yet. A modified form of multiplicative $\sin\theta$ term is needed, but a numerically reliable modification is not yet known. 

The widths of the peaks are determined by the finite radial and angular resolution. The radial resolution is set by the maximum $q$ value used in the calculation of the correlation function, which in this case is set by the distance from the centre of the diffraction pattern to the detector edge. The angular resolution of the PADF is set by the maximum value of $l$ used in the calculation, which was $l_{max} = 30$. 

There is angular structure that is weaker than the principal peaks observable at angles not expected from the ideal structure. These are artefacts created by truncating the basis sets used, which are analogous to Fourier truncation artefacts in signal processes caused by the truncated Fourier series expansion. The truncation artefacts can be reduced by increasing the angular and radial resolution, but cannot be removed entirely due to resolutions at which experiments will reasonably converge.

     
\section{Access to $\textit{pypadf}$}   
\;\\  
The \textit{pypadf} package can be downloaded from https://github.com/amartinrmit/pypadf and is distributed under the GNU Lesser General Public License (LGPL,
Version 3; https://www.gnu.org/licenses/lgpl-3.0) . The \textit{pypadf} package is written in  \textit{Python 3} and requires the following packages: \textit{numpy}, \textit{scipy}, \textit{matplotlib}, \textit{numba}, the python imaging library (PIL) and \textit{h5py}. There is a tutorial with a detailed explanation of each script and the configuration files for the model hexagonal sample shown here are included with the code. The readme.md file contains instructions for install and tutorial for creating diffraction of hexagon structures, correlating them, and calculating the padf. A list of possible parameters can be found for the scripts can be found with the --help commandline argument.

     
\section{Conclusion and Future Work}
\;\\  
The \textit{pypadf} package has been presented that can compute the PADF from fluctuation scattering diffraction data. The package includes scripts that can simulate diffraction patterns, compute angular correlation functions, modify angular correlation functions, compute the PADF and finally plot the results. The analysis assumes kinematic scattering approximations and that each diffraction pattern is of a sample in a random orientation or a statistically independent region of a bulk disordered sample. We expect the code to be useful for probing local 3D structures in disordered materials probed with x-ray and electron beams. Scanning diffraction and serial diffraction experiments are well-established data collection methods with electron microscopes, synchrotrons and x-ray free-electron laser facilities. We expect that many existing fluctuation datasets are suitable for PADF analysis and that many facilities already have the capability to measure these datasets. Further work is still required to understand convergence of the correlation functions and how to reduce numerical artefacts in linear transformations, interpolations and matrix inversions that are used. For electron diffraction calculations, the effect of dynamical diffraction is yet to be investigated.


\appendix
\section{Diffraction Scripts}
\label{Appendix:Diffraction}
\subsection{The input data: far-field diffraction patterns}

\subsubsection{Diffraction: mathematical and numerical details}
\;\\
The input data for the correlation calculation are a set of diffraction patterns $I(\textbf{q})$, where $\textbf{q} = \textbf{q}_s - \textbf{q}_0$ is the difference between the scattering $\textbf{q}_s$ and incident $\textbf{q}_0$ wave vectors. The kinematic diffraction approximation (single, elastic scattering) is assumed. 
The intensity for uniform incident pulse is given by
\begin{equation}
I(\textbf{q}) = r_e^2 P(\textbf{q}) \frac{ \phi_0} {A^2} d\Omega |F(\textbf{q})|^2 \;,
\label{eq:Iq}
\end{equation}
where $r_e$ is the classical electron radius, $P(\textbf{q})$ is a polarisation factor, $d\Omega$ is the solid angle of a pixel. $A$ is the beam area and $F(\textbf{q})$ is the molecular scattering factor, which is calculated from the atomic positions.

The scattering angle of the $i^{th}$ pixel is defined as 
\begin{equation}
\theta_{i} = \frac{1}{2}\arctan\left( \frac{r_i}{z} \right) \;,
\label{eq:thetai}
\end{equation}
where $r_i$ is the radial distance of the pixel centre from the beam center and $z$ is the sample to detector distance. In the \textit{pypadf} package, the magnitude of the vector $\textbf{q}_i$ associated with the $i^{th}$ pixel is defined to be
%
\begin{align}
q_i &\equiv |\textbf{q}_i| \nonumber \\
 &= \frac{2}{\lambda} \sin\theta_i \;.
\label{eq:qi}
\end{align}
We note that this definition uses a convection common in electron scattering applications and differs from the usual convention in x-ray diffraction by a factor of $2\pi$. In x-ray sciences $q_{x-ray} = 2 \pi |\textbf{q}_i| = \frac{4 \pi}{\lambda} \sin\theta_i$. 

A consequence of Ewald sphere curvature, expressed by Eqs (\ref{eq:thetai}) and (\ref{eq:qi}), is that pixels of a uniform width do not generate uniform sampling of $\theta_i$ and $q_i \equiv |\textbf{q}_i|$. To facilitate the calculation of angular correlations, the diffraction pattern can be represented in polar coordinates $I(q,\theta)$, where $q$ is a vector magnitude of $\textbf{q}$ and $\theta$ is an angle around the beam axis. Interpolation of the diffraction data into the polar representation is implemented using the \textit{map\_coordinates} function from the \textit{scipy.ndimage} module~\cite{Virtanen2020}, and in the same step the irregular sampling of the radial $q$ coordinate, originating from the Ewald sphere, is converted to regular $q$-sampling.

A second effect of the Ewald sphere is that the solid-angle subtended by a pixel reduces if the pixel is located further away from the beam center. The reduction in solid angle is approximately given by  $d\Omega_i = d\Omega_0 \cos \theta_{i}$, where $d\Omega_0$ is the solid angle of a pixel at the beam centre. There is an option to apply this approximation in diffraction pattern simulations and the effect becomes significant for wide-angle diffraction.

%
%
%

\subsubsection{\textit{diffract.py} : computing test diffraction data}

The script \textit{diffract.py} can be used to compute basic diffraction data for testing the correlation and PADF scripts. This script computes a diffraction pattern of a randomly orientated molecule from atomic coordinates given in a Protein Data Bank file (.pdb). It does not use any unit cell or crystal lattice information and, hence, only simulates continuous diffraction. The atomic scattering factors are taken from Waasmeier and Kirfel~\cite{Waasmaier1995} and data from Henke et al.~\cite{Henke1993} are used for wavelength dependent corrections. A square detector is assumed and its distance from the sample, pixel width and number of pixels along a side length can be varied. 

\subsubsection{\textit{diffract\_and\_correlate.py} : testing large data amounts}
 \;\\

 When simulating large diffraction datasets, it can be impracticable to store every diffraction pattern prior to calculating the correlation function. The script $\textit{diffract\_and\_correlate.py}$ simulates diffraction patterns and correlates those patterns on the fly. Only a small number of diffraction patterns (a chunk) are created at any one time, then they are correlated and deleted, and the cycle is repeated. A single atomic structure can be used, which can be randomly rotated and translated with periodic boundary conditions to generate the diffraction patterns. This script takes the same parameters as \textit{diffract.py} and \textit{difftocorr.py}.



\ack{Acknowledgements}

We acknowledge the funding support from the Australian Research Council Discovery Project grant
(DP190103027).

\pagebreak

%
%

\pagebreak
%
%

\pagebreak
%
%

\begin{table}
\label{table:hexpeaks}
\caption{Ratio of the peak heights extracted from the simulation. The labels A, B and C are defined in Fig. \ref{fig:hexexample}. $R_{A/B}$ stands for the ratio of the height of peak A to the height of peak B. }
\begin{tabular}{ccc}      
 Peak ratio    & Ideal value & Simulated value           \\
\hline
$R_{B/A}$     & 1       &   0.95  \\
$R_{C/A}$     & 0.25    &   0.27     \\
\end{tabular}
\end{table}
\pagebreak

\referencelist[pypadf]

\end{document}